\documentclass[sigconf,authorversion,nonacm,screen]{acmart}
\AtBeginDocument{%
  \providecommand\BibTeX{{%
    \normalfont B\kern-0.5em{\scshape i\kern-0.25em b}\kern-0.8em\TeX}}}

\usepackage{soul}
\usepackage{balance}
\usepackage{caption}
\usepackage{subcaption}
\usepackage{tabularx}
\usepackage{makecell}

\usepackage{multirow}
\usepackage{CJKutf8}
\usepackage{makecell}
\usepackage{acmart-taps}

\copyrightyear{2024}
\acmYear{2024}
\setcopyright{acmlicensed}\acmConference[CHI '24]{Proceedings of the CHI Conference on Human Factors in Computing Systems}{May 11--16, 2024}{Honolulu, HI, USA}
\acmBooktitle{Proceedings of the CHI Conference on Human Factors in Computing Systems (CHI '24), May 11--16, 2024, Honolulu, HI, USA}
\acmDOI{10.1145/3613904.3642273}
\acmISBN{979-8-4007-0330-0/24/05}

\begin{document}
\begin{CJK}{UTF8}{ipxm}

\title[Play Across Boundaries]{Play Across Boundaries:\\Exploring Cross-Cultural Maldaimonic Game Experiences}

\author{Katie Seaborn}
\email{seaborn.k.aa@m.titech.ac.jp}
\orcid{0000-0002-7812-9096}
\affiliation{
  \institution{Tokyo Institute of Technology}
  \city{Tokyo}
  \country{Japan}
}

\author{Satoru Iseya}
\orcid{0000-0002-0228-4671}
\email{iseya.s.aa@m.titech.ac.jp}
\affiliation{
  \institution{Tokyo Institute of Technology}
  \city{Tokyo}
  \country{Japan}
}

\author{Shun Hidaka}
\orcid{0000-0003-3132-2099}
\email{hidaka.s.ad@m.titech.ac.jp}
\affiliation{
  \institution{Tokyo Institute of Technology}
  \city{Tokyo}
  \country{Japan}
}

\author{Sota Kobuki}
\orcid{0000-0003-0531-7579}
\email{kobuki.s.aa@m.titech.ac.jp}
\affiliation{
  \institution{Tokyo Institute of Technology}
  \city{Tokyo}
  \country{Japan}
}

\author{Shruti Chandra}
\email{shruti.chandra@uwaterloo.ca}
\orcid{0000-0003-2042-8875}
\affiliation{
  \institution{University of Waterloo}
  \city{Toronto}
  \country{Canada}
}
\affiliation{
  \institution{IRFI, Tokyo Institute of Technology}
  \city{Tokyo}
  \country{Japan}
}

\renewcommand{\shortauthors}{Seaborn et al.}

\begin{abstract}
Maldaimonic game experiences occur when people engage in personally fulfilling play through egocentric, destructive, and/or exploitative acts. Initial qualitative work verified this orientation and experiential construct for English-speaking Westerners. In this comparative mixed methods study, we explored whether and how maldaimonic game experiences and orientations play out in Japan, an Eastern gaming capital that may have cultural values incongruous with the Western philosophical basis underlying maldaimonia. We present findings anchored to the initial frameworks on maldaimonia in game experiences that show little divergence between the Japanese and US cohorts. We also extend the qualitative findings with quantitative measures on affect, player experience, and the related constructs of hedonia and eudaimonia. We confirm this novel construct for Japan and set the stage for scale development.
\end{abstract}




\maketitle

\section{Introduction}

The field of human-computer interaction (HCI) has long recognized the centrality of user experience (UX) in the design and study of interactive experiences, including video games~\cite{anderson_seductive_2011,hassenzahl_user_2008,hassenzahl_user_2006,korhonen_understanding_2009}. While imprecise and far-reaching~\cite{forlizzi_understanding_2004}, UX can be defined as a person's overall subjective experience, which may be more complex than merely ``positive'' or ``pleasurable''~\cite{hassenzahl_user_2006,hassenzahl_user_2008,hassenzahl_user_2006,korhonen_understanding_2009,wright_making_2004}. A useful operationalization has been found in the concepts of \emph{hedonia} as pleasure-seeking and \emph{eudaimonia} as meaning-making~\cite{huta_pursuing_2013,deci_hedonia_2008}, entwined constructs that hail from ancient Greek philosophy. While most work in the context of media and interactive experiences has focused on hedonia~\cite{diefenbach_hedonic_2014, hancock_hedonomics_2005, hassenzahl_effect_2001, helander_hedonomicsaffective_2002}, a growing body of work has explored eudaimonia~\cite{deterding_eudaimonic_2014, mekler_momentary_2016, muller_facets_2015, seaborn_evaluating_2016,tamborini_defining_2010,bartsch_appreciation_2017, oliver_entertainment_2011, wirth_beyond_2012,cole_thinking_2021, daneels_eudaimonic_2021,seaborn_eudaimonia_2020}, indicating a new focus on meaning and meaningful engagement.

Recently, the notion of \emph{maldaimonia} was proposed as a subversive spin on eudaimonia: an orientation towards self-expression that is egocentric, exploitative, and/or destructive \cite{waterman_toward_2021}. However, as a novel concept, it is not well understood in practice and has no construct validation or measure yet.
Only one preliminary study has explored maldaimonia empirically---in the context of games~\cite{seaborn_meaningful_2023}. 
Similar concepts exist, from \emph{transgressive play} \cite{aarseth2014fought,jorgensen_transgression_2019, mortensen_paradox_2020, sidhu2023benevolent} to \emph{dark play} \cite{mortensen_dark_2015} to \emph{cruel play} \cite{sutton-smith_ambiguity_2001}, and \emph{dark participation} \cite{kowert_dark_2020,kowert_toxicity_2022,tang_mens_2016}. What seems to set the notion of maldaimonic game UX apart is its dual nature: \emph{self-focused} or \emph{social}, having been found in single-player contexts \emph{and} with known others~\cite{seaborn_meaningful_2023}. In kind, we approach maldaimonia as an \emph{orientational frame} characterized by \emph{positive}, \emph{meaningful}, \emph{self-realizing}, and \emph{self-actualizing} play.

Maldaimonia is a concept created within the West and driven by Western cultural roots. As such, it is potentially Western-centric or -centr\emph{ized}, limiting its generalizability and maintaining the WEIRD (Western Educated Industrial Rich Democratic) bias~\cite{henrich_weirdest_2010} in and beyond HCI research~\cite{linxen_how_2021}. A clear next step was to explore its applicability---and potentially its variations---in other cultures. One option was Japan, an Eastern nation known for its lively gaming culture. Japan is home not only to heavyweights like Nintendo and Sony, but also a large playership among the general population. Statistics consistently place Japan at the top of user penetration lists, with 58\% user penetration in 2022\footnote{\url{https://www.statista.com/outlook/amo/media/games/japan}}; in comparison, the US ranks less than 45\%\footnote{\url{https://www.statista.com/topics/3070/us-gamers}}. Japan and the US have also long been compared at a grand cultural scale through Hofstede's cultural dimensions~\cite{hofstede2001culture}. 
Japan thus offers an ideal context for exploring the Western notion of maldaimonia in games and comparing the findings from the preliminary study~\cite{seaborn_meaningful_2023} on US players.

In this work, we conducted an online critical incident survey on Japanese people's experiences of maldaimonia in games, comparing to the English data set captured by \citet{seaborn_meaningful_2023}, who used the procedures of \citet{muller_facets_2015} and \citet{mekler_momentary_2016} for eudaimonic UX. Our goal was to find further evidence of maldaimonic game UX and establish its cross-cultural viability as an experiential construct and orientation. Our overarching research question (RQ) was: \textbf{\emph{Is maldaimonic game UX a cross-cultural phenomenon?}} Specifically, we sought to answer: \emph{RQ1: Do Japanese players have similar maldaimonic game experiences as US players?} and \emph{RQ2: Does maldaimonia map onto relevant theoretical, orientational, experiential, and game UX factors, notably hedonia and eudaimonia, player experience, and affect?} 
We found that players in Japan have maldaimonic game experiences and orientations very similar to those found among players in the English-speaking West, with caveats. Our contributions are: (i) further empirical support for maldaimonic game UX as an experiential construct and orientation; (ii) cross-cultural viability of the concept, despite its Hellenic roots; (iii) enriched findings through validated measures of relevant factors; and (iv) implications for further theorizing and study. This work adds a slightly different take to research on ``dark'' or ``transgressive'' play, brings in a crucial multicultural perspective, and sets the stage for measure development.

\section{Theoretical Background}
We begin by outlining the theory behind maldaimonia and drawing together adjacent concepts in the well-being and game UX literatures. We present an overview of how these concepts are connected and differentiated in \autoref{tab:concepts}.

\begin{table*}
\caption{Overview of concepts related to maldaimonia.}
\label{tab:concepts}
\begin{tabularx}{\textwidth}{p{70pt}p{150pt}p{60pt}p{45pt}p{135pt}}
\hline
\makecell{Concept} & Definition & Distinction & Ethics & Source \\
\hline
\makecell{Maldaimonia} & Personal fulfillment that is egocentric, exploitative, and/or destructive & Orientations and experiences & Negative & \citet{waterman_toward_2021} \\
\makecell{Eudaimonia} & Personal fulfillment that is allocentric (both egocentric and altruistic), reciprocal, and/or constructive, with a view to future happiness & Orientations and experiences  & Positive & \citet{deci_hedonia_2008,huta_pursuing_2010,huta_eudaimonia_2013} \\
\makecell{Hedonia} & Pleasure-seeking and pain-avoidance & Orientations and experiences & Ambiguous & \citet{deci_hedonia_2008,huta_pursuing_2010,huta_eudaimonia_2013} \\
\makecell{Transgressive play} & Acts of player rebellion against the intended experience of the game that may be novel, subversive, or offensive & Behaviour & Ambiguous & \citet{aarseth2014fought,sidhu2023benevolent} \\
\makecell{Dark play} & Aspects of the game intentionally designed to provoke, shock, and/or disturb the player & Stimuli & Ambiguous & \citet{mortensen_dark_2015}, \citet{mortensen_paradox_2020} \\
\makecell{Cruel play,\\dark participation} & Player actions during gameplay that are harmful to others, often on purpose & Behaviours and experiences & Negative & \citet{sutton-smith_ambiguity_2001,kowert_dark_2020,kowert_toxicity_2022,tang_mens_2016} \\
\hline
\end{tabularx}
\end{table*}

\subsection{From Eudaimonia to Maldaimonia}
Waterman based the concept of maldaimonia on \emph{eudaimonia}. Originating from Aristotle's \emph{Nichomachean Ethics} in the 4th century BCE~\cite{waterman_toward_2021}, eudaimonia refers to an inclination towards future happiness and present flourishing~\cite{deci_hedonia_2008,huta_pursuing_2013,huta_eudaimonia_2013,waterman_personal_1990,waterman_reconsidering_2008,waterman_two_1993}. This \emph{orientational} and \emph{experiential construct} represents the pursuit of meaningful engagement and self-expression. Maldaimonia emerged from a critical examination of classifying \emph{ethically negative} orientations, actions, and experiences as ``eudaimonic.'' Waterman argues that activities characterized by egocentrism, destructiveness, exploitation, and harm cannot be considered virtuous, good, or desirable, i.e., \emph{ethically positive}, even if maldaimonia is a type of ``flourishing''~\cite{haybron_philosophical_2016}. As \citet[p. 2]{seaborn_meaningful_2023} ponder, ``If eudaimonia is about \emph{virtue}, could maldaimonia be about \emph{vice}?'' Waterman offers four characteristics of maldaimonia: (i) attaching positive value to self-centered, destructive, and exploitative actions; (ii) forming a basis for personal identity; (iii) striving for excellence or mastery in these actions; (iv) and aligning these actions with personal expressiveness. Consequently, in theory, maldaimonia is linked to eudaimonia through the pursuit of \emph{affectively positive} and \emph{meaningful} engagement. Indeed, this is what \citet{seaborn_meaningful_2023} found for English-speaking US gamers. As such, \textbf{we hypothesized that the baseline characteristics of eudaimonia as positive (H1) and meaningful (H2) would be found for maldaimonia}, as well.

Still, there are two notable and interconnected omissions from Waterman's criteria. One is \emph{hedonia}, which is always significantly correlated with eudaimonia \cite{huta_pursuing_2010,huta_eudaimonia_2013} but has no theorized equivalent in the conceptualization of maldaimonia. 
Perhaps hedonia is ethically neutral or \emph{ambiguous}, without a counterpart for maldaimonia. We must consider, for instance, the value of confronting painful truths and avoiding over-indulgence in life's pleasures. Still, \textbf{we hypothesized that, if maldaimonia is the wicked twin of eudaimonia, hedonia should be present, too (H3)}.

The other missing element is \emph{long-term meaning potential}, which has been associated with eudaimonia generally~\cite{huta_eudaimonia_2013,huta_pursuing_2010} and eudaimonic UX specifically~\cite{mekler_momentary_2016}.
Where hedonia relates to short-term 
goals, eudaimonia has a longitudinal component: not only meaningful engagement in the now, but also an orientation towards future happiness~\cite{waterman_reconsidering_2008, waterman_two_1993,huta_pursuing_2010, huta_pursuing_2013}. 
\textbf{We thus hypothesized that measures of eudaimonia as a function of maldaimonia would relate to long-term meaning potential but not hedonia (H4)}.

\subsection{Maldaimonia-Adjacent Concepts in Games and Play}
The notions of negative, transgressive, and subversive play are not new. On the ethically \emph{negative} front, Sutton-Smith long ago recognized that even children engage in \emph{cruel play}~\cite{sutton-smith_ambiguity_2001}. In games, \emph{dark participation}~\cite{kowert_dark_2020,kowert_toxicity_2022} encompasses behaviours and experiences of toxicity, trolling, bullying, deviancy, and harassment, where player needs are satisfied through harming others. ``Toxic gamer cultures'' represent a brand of experiences within gaming communities as well as within gameplay. This phenomenon is characterized by exclusion and hostility, if not outright harassment, via anonymous actors who can shirk responsibility and hide behind anonymity~\cite{kowert_dark_2020,tang_mens_2016}. Such interactions are not necessarily tied to the game itself, i.e., people doing bad things to each other because these spaces allow for it to happen. That people may engage in solo play or agree with others to have fun ``being bad'' in games is less explored~\cite{seaborn_meaningful_2023}. Moreover, an \emph{orientational} perspective that speaks to psychological needs, states, and cognitions, especially the meaning behind such behaviours, is lacking. Maldaimonia could fill this gap, bringing clarity to the
difference between those agreeing to be bad for fun and a bad actor forcing an undesirable experience on others.

\emph{Transgressive play} refers to several related ideas that broadly capture acts of subversion (ethically \emph{ambiguous}) or transgression (ethically \emph{negative}) by the player or embedded in the gameplay~\cite{jorgensen_transgression_2019}. \citet{aarseth2014fought} presents the concept as player behaviour that reflects accidental or willful disregard of the ``law'' of the game, describing scenarios of almost whimsical and unintended rebellion. \citet{sidhu2023benevolent} proposed \emph{benevolent} transgressive play as \emph{creative} acts by the player that buck the rules of the game. On the flip side, transgressive play and \emph{dark play}~\cite{mortensen_dark_2015,mortensen_paradox_2020} refer to cases where game designers have orchestrated an experience that subjects the player to play meant to provoke, disturb, and/or shock. The purpose and outcomes run the moral gamut and so are \emph{ambiguous}. Maldaimonia relates to but is distinct from these concepts. It is ethically negative, the moral opposite of eudaimonia. It is not a stimulus or a behaviour, although it may be triggered by a stimulus or be observed through player behaviour. It is an orientation that may or may not reflect a rebellious spirit. Importantly, it may further our understanding of transgressive game stimuli and player behaviours from a psychological angle~\cite{waterman_toward_2021,seaborn_meaningful_2023}. \textbf{We thus expect and hypothesize that maldaimonia and the experiences it occasions will be associated with good player experience, satisfying player functional, psychological, and social needs (H5)}.

Maldaimonic acts are not unprecedented in games.
Cheating is one form of maldaimonia posed by \citet{waterman_toward_2021} with a long history in game studies \cite{consalvo_cheating_2009}. For instance, \citet{passmore_cheating_2020} explored the affective and psychological benefits to cheating within solo gaming experiences. While many players felt frustration and shame, a comparative portion felt joy, agency, relief, surprise, and excitement. Destructive acts have been less explored, even though they are common in games: consider \emph{Catlateral Damage} (2014) or any game with destruction built-in, like \emph{Gunstar Heroes} (1993). \citet{pimentel_your_2020}, for instance, explored how people reacted to destroying 
NPCs customized to represent positive or negative self-concepts. They found that destroying these characters increased negative affect when the player identified with them more. \citet{grizzard_being_2014} explored whether games could be designed to induce guilt, comparing the experiences of people who took on the role of a terrorist, a potentially maldaimonic orientation, or a UN soldier. They found that guilt-inducing experiences not only occurred but increased moral sensitivity. 

What this body of work and the preliminary findings~\cite{seaborn_meaningful_2023} suggest is that maldaimonic orientations and experiences exist across a variety of games, within and outside of multiplayer contexts, and have implications for emotional, social, and ethical engagement. Still, we do not know its extent and presence in general or across cultural contexts. Given the dearth of empirical work on maldaimonia, \textbf{we hypothesized that there would be no cross-cultural differences (H6)}, a baseline assumption to be confirmed or refuted~\cite{ember2009cross}. In this work, we compared the Western context of America, which, like maldaimonia, has roots in Hellenic culture, to the Eastern context of Japan: two different cultural contexts with a shared love of games that, when juxtaposed, could reveal whether and where differences in maldaimonic game UX occur.

\section{Methods}
We carried out an online survey to capture the maldaimonic experiences of Japanese players, replicating the procedure of \citet{seaborn_meaningful_2023} as registered on OSF\footnote{Registered before data collection on on July 5\textsuperscript{th}, 2022: \url{https://osf.io/7cvs9}}. This study was approved by the university ethics committee 
on August 9\textsuperscript{th}, 2022 (\#2022153).

We used the critical incident method~\cite{woolsey_critical_1986}, also called the critical incident technique (CIT)\footnote{\url{https://www.nngroup.com/articles/critical-incident-technique/}}, which is a systematic qualitative approach to collecting self-reports on significant \emph{incidents}---events, occurrences, and experiences---from individuals. Participants are guided with clear instructions and prompts (refer to~\ref{sec:procedure}) to think about and describe a single incident that they deem critical in some way. Here, ``critical'' is a property of the experience that the participant feels had a substantial impact on their attitudes or behaviour in relation to the outcome of a given activity. In this research, the experience was an instance of gameplay and the critical incident involved maldaimonia affecting the nature or outcome of the play experience. Given the nature of maldaimonia, we expect such experiences to be memorable and impactful. This method has long been used in HCI and related areas~\cite{woolsey_critical_1986,hassenzahl_experience_2010,mekler_momentary_2016,muller_facets_2015}.

Except where noted, we translated the materials and instruments ourselves, following the ITC Guidelines for Translating and Adapting Tests~\cite{international_test_commission_itc_2018}. This meant: being knowledgeable about the socio-linguistic context, i.e., Japanese, and the content, i.e., maldaimonia; forward-translation and backward-translation to ensure retention of meaning; double-checking with native speakers and running pilot tests; using conventional response formats; sampling an appropriate population; and providing evidence of reliability.

\subsection{Participants}
Responses were collected via Yahoo! Crowdsourcing, an anonymous online recruitment service that guarantees unique identities and demographics using account verification\footnote{\url{https://crowdsourcing.yahoo.co.jp}}. There were 59 Japanese adults aged 21+.
This sample size falls within the range appropriate for the critical incident method (e.g., N=45~\cite{partala_understanding_2012}). 13 (22\%) women and 46 (78\%) men participated, with none of anther gender identity. Most (58 or 98.3\%) were East Asian, i.e., Japanese, and one (1.7\%) was mixed East Asian and White. Most had at least a high school education (58 or 98.3\%). They were remunerated with 300 yen (USD $\sim$\$2). A verification code was used to ensure that participants completed the full questionnaire. The study was approved by the research institution's ethics board on 9\textsuperscript{th} August 2022.

\subsection{Procedure} \label{sec:procedure}
A link to a Google Form containing the survey was provided on Yahoo! Crowdsourcing. The consent form was presented on the first page with an explanation of the study and a definition of maldaimonia to filter in participants:

\aptLtoX[graphic=no,type=html]{
\begin{quote}\ 
マルダイモニアと は、「自己中心的、破壊的、搾取的な行為を通じて、楽しさや自己実現を感じることを指しま す。ゲームでは、敵を殺す、人のものを盗む、街を破壊するなどがこれにあたります。これら は一例であり、他にもあるかもしれません。」 \hfill\break
``Maldaimonia'' refers to feelings of enjoyment and self-fulfillment through egocentric, destructive, and/or exploitative acts. In games, this could include, for example, killing enemies, stealing from others, and destroying cities. These are just examples; there may be more.
\end{quote}
}
{\begin{quote}マルダイモニアと は、「自己中心的、破壊的、搾取的な行為を通じて、楽しさや自己実現を感じることを指しま す。ゲームでは、敵を殺す、人のものを盗む、街を破壊するなどがこれにあたります。これら は一例であり、他にもあるかもしれません。」 \\
``Maldaimonia'' refers to feelings of enjoyment and self-fulfillment through egocentric, destructive, and/or exploitative acts. In games, this could include, for example, killing enemies, stealing from others, and destroying cities. These are just examples; there may be more. 
\end{quote}}

Participants then submitted their report and answered the rest of the questionnaire, with demographics at the end. After submitting their answers, they were given a code for Yahoo! Crowdsourcing to receive compensation. The survey took about $\sim$15 minutes.

\subsection{Qualitative Instrument}
We asked participants to report a critical incident on a single maldaimonic experience using a series of prompts and open-ended questions, as in previous research~\cite{seaborn_meaningful_2023,mekler_momentary_2016}. As per the method~\cite{woolsey_critical_1986}, we included all the prompts needed to elicit details about the experience, replicating the preliminary study~\cite{seaborn_meaningful_2023}. Respondents were prompted to bring to mind the experience, priming them to think about it while completing the entire survey:

\aptLtoX[graphic=no,type=html]{
\begin{quote}\ 
あなたがこれまでにゲームで体験した「マルダイモニア」を1つだけ思い浮かべてください。あなたにとって「マルダイモニア」と思う体験を思い浮かべてください。スマートフォン、パソコン、家庭用ゲーム機など、ゲームの種類やプラットフォームは問いません。ソロでもマルチでもOK。あなたが自発的にマルダイモニア的な行動を選んだものでも、他の誰かがしたものでも、ゲームプレイで要求されたものでもOKです。
\hfill\break
Bring to mind a single "maldaimonic" experience you've had in a game. Think of "maldaimonia" in whatever way that makes sense to you. You can choose any type of game on any platform, including games played on a smartphone, a computer, a console, etc. You can share a solo or multiplayer experience. It can be one where you chose to act in a maldaimonic way or someone else did, or it was required by the gameplay.
\end{quote}
}
{\begin{quote}
あなたがこれまでにゲームで体験した「マルダイモニア」を1つだけ思い浮かべてください。あなたにとって「マルダイモニア」と思う体験を思い浮かべてください。スマートフォン、パソコン、家庭用ゲーム機など、ゲームの種類やプラットフォームは問いません。ソロでもマルチでもOK。あなたが自発的にマルダイモニア的な行動を選んだものでも、他の誰かがしたものでも、ゲームプレイで要求されたものでもOKです。\\
Bring to mind a single "maldaimonic" experience you've had in a game. Think of "maldaimonia" in whatever way that makes sense to you. You can choose any type of game on any platform, including games played on a smartphone, a computer, a console, etc. You can share a solo or multiplayer experience. It can be one where you chose to act in a maldaimonic way or someone else did, or it was required by the gameplay.
\end{quote}
}

Next, we asked which platform they used, e.g., PlayStation, their motivation for playing the game, and who they played with, if anyone.
The last question was important for distinguishing maldaimonia as a positive solo or social activity from dark participation as a negative social activity~\cite{kowert_dark_2020,seaborn_meaningful_2023}. They were then asked:
\aptLtoX[graphic=no,type=html]{
\begin{quote}\ 
あなたがゲーム内で経験した『マルダイモニック』な体験を記述してください。ゲーム内での自己中心的、破壊的、搾取的な行為を通じて、どのような楽しさや自己実現の感情を抱いたかに着目してください。なるべく詳しくお願いします
\hfill\break
Describe the ``maldaimonic'' experience you had in the game. Focus on how your experience involved feelings of enjoyment and self-fulfillment through egocentric, destructive, and/or exploitative acts in the game. Please be as detailed as possible.
\end{quote}
}
{\begin{quote}
あなたがゲーム内で経験した『マルダイモニック』な体験を記述してください。ゲーム内での自己中心的、破壊的、搾取的な行為を通じて、どのような楽しさや自己実現の感情を抱いたかに着目してください。なるべく詳しくお願いします。\\
Describe the ``maldaimonic'' experience you had in the game. Focus on how your experience involved feelings of enjoyment and self-fulfillment through egocentric, destructive, and/or exploitative acts in the game. Please be as detailed as possible.
\end{quote}}

They were also asked to describe what motivated their behaviour, i.e., their maldaimonic orientation or drivers behind the maldaimonic experience, as follows:

\aptLtoX[graphic=no,type=html]{
\begin{quote}\ 
マルダイモニックな方法でゲームをプレイすることは、その動機は何ですか?
\hfill\break
What was the motivation behind playing the game in a maldaimonic way?
\end{quote}
}
{\begin{quote}
マルダイモニックな方法でゲームをプレイすることは、その動機は何ですか？\\
What was the motivation behind playing the game in a maldaimonic way?
\end{quote}
}

\subsection{Quantitative Instrument}

All instructions referenced the single maldaimonic game experience rather than gaming experiences in general.

\subsubsection{Positive and Negative Affect (H1)}
As \citet{waterman_toward_2021} outlined, positive affect is important for the construct of maldaimonia. At the same time, maldaimonia is premised on negative engagements and orientations. We used the PANAS-X \cite{watson_development_1988}, a widely used and validated self-report instrument for capturing affective states. We could not predict what kind of affective qualitative maldaimonic experiences might have, so we included all terms from the basic negative, basic positive, and other affective scales. These were presented in random order on 5-point Likert scales operationalized as a 0--4 scale (0=強くそう思う、4=強くそう思わない or 0=strongly agree, 4=strongly disagree).

\subsubsection{Hedonic and Eudaimonic Motivations (H2, H3)} We used the Japanese translation~\cite{asano2021psychometric} of the Hedonic and Eudaimonic Motivations for Activities (HEMA) scale \cite{huta2014eudaimonia} to evaluate meaning, in the absence of a scale for maldaimonia (H2), as well as the role of hedonia (H3). This 9-item cross-culturally validated scale assesses orientations to engage in an activity through a 1\phantom{ }(全くあてはまらない or not at all) to 7\phantom{ }(非常にあてはまる or very much) response scale. Example items are:\phantom{ }くつろぎを追求すること (seeking comfort) and\phantom{ }自分の信念に従った行動を追求すること (seeking to do what you believe in). We excluded the extra items in \citet{asano2021psychometric} so as to directly compare the English and US data sets. Theoretically, there should be high levels of eudaimonia and the presence of hedonia, given eudaimonia's theorized connection to maldaimonia \cite{waterman_toward_2021} and how hedonia overlaps with eudaimonia \cite{huta_eudaimonia_2013}.

\subsubsection{Player Experience (H2, H5)}
We used the Player Experience Inventory (PXI) \cite{abeele_development_2020} for measuring player experience, which has been validated with over 64 games and 529 players. The instrument was built around two theories: Means-Ends theory \cite{gutman_means-end_1982} and the Mechanics, Dynamics, and Aesthetics (MDA) model \cite{hunicke_mda_2004}. We used the Psychosocial Consequences (H2) and Functional Consequences (H5) subscales. The first is expected to link to eudaimonia and potentially maldaimonia given conceptual overlap of the constructs with respect to meaningfulness: Curiosity, Mastery, Immersion, and Autonomy. The second relates to matters of game usability and control, which could inform positive or negative experiences, e.g., bad usability and poor control. A seven-point Likert scale of 0±3 was used as per the instructions \cite{abeele_development_2020}.

\subsubsection{Long-term Meaning Potential (H4)}
We evaluated the long-term meaning potential of the maldaimonic experience using a one-item scale. Previous work \cite{kim_pleasure_2014, mekler_momentary_2016,seaborn_meaningful_2023} found that pleasure was preferred sooner rather than later, while the opposite was true for meaningful experiences, which are enriched over time~\cite{kim_pleasure_2014}. These are correlates of hedonia and eudaimonia and may help make sense of maldaimonia in terms of temporal distance. We used a 7-point importance scale (1=全く重要でない、7=とても重要である or 1=not important at all, 7=very important). The item was:

\aptLtoX[graphic=no,type=html]{
\begin{quote}\ 
1年後の自分の人生を考えた場合、この体験はどの程度重要だと思いますか?
\hfill\break
If you consider your life one year from now, how important will you find this experience?
\end{quote}
}
{\begin{quote}
1年後の自分の人生を考えた場合、この体験はどの程度重要だと思いますか？\\
If you consider your life one year from now, how important will you find this experience?
\end{quote}
}

\subsection{Data Analysis}
We carried out complementary qualitative and quantitative analyses. We provide our data set\footnote{\url{https://bit.ly/maldaimoniajp}} on the Japanese cohort (N=59) for open science. We used the open data set\footnote{\url{https://osf.io/cpfzy}} on the US cohort (N=51) provided by \citet{seaborn_meaningful_2023} in all comparative analyses with our Japanese cohort for cross-cultural verification (H6).

\subsubsection{Qualitative Analysis}
We used hybrid thematic analysis \cite{proudfoot_inductivedeductive_2023}, combining the deductive application of existing themes with inductive theme development.

For the motivations data, we applied the maldaimonic play motivations framework from \citet{seaborn_meaningful_2023} but expanded upon it in two ways. First, we developed higher-order themes---Passivity, Provocation, Mentalization, and Affective Drivers---in pursuit of a framework that could be used for future comparisons of maldaimonia to other concepts (\autoref{tab:motivthemes}). Second, we applied Waterman's four criteria, which should also describe motivations, being grounded in one's orientation~\cite{waterman_toward_2021}. For the experience data, we used the maldaimonic game experiences framework crafted by \citet{seaborn_meaningful_2023}, without modifications (Table~\ref{tab:malthemes}).

We used inductive theme development to account for culturally-sensitive ideas. The first author considered how aspects of the accounts unrelated to the deductive themes may be explained by facets of Japanese culture. Notably, while advanced in Japanese and having seven years of experience conducting research in the Japanese context, the first author is not Japanese, thus limiting their ability to identify new themes with fluency. First, they completed the deductive analysis, which led to the identification of new themes, as per Braun and Clarke~\cite{braun_using_2006}. One new theme---Influence---was constructed for the maldaimonic play motivations framework. They also found instances of this new theme in the US data set.

For the deductive analyses, four raters divided and separately coded all experience accounts and motivation data. Three were native Japanese speakers and one (the first author) was advanced in Japanese. Percentage agreement was used to evaluate inter-rater reliability (IRR) between pairs of raters given the relatively small samples and large number of possible themes and codes~\cite{birkimer1979back}. 
When disagreements occurred, the raters discussed and re-coded.

We analyzed the extent to which reports of experiences and motivations mapped onto Waterman's four criteria of maldaimonia, aiming to determine how well actual accounts matched maldaimonia as theorized~\cite{waterman_toward_2021}. We generated counts and percentages for experiences, motivations, and both together.

We used Chi-square tests to evaluate the differences in relative counts for each theme by country.

\begin{table*}
\caption{Maldaimonic play motivations framework, an extended version from \citet{seaborn_meaningful_2023}. Criteria refer to the four criteria proposed by Waterman~\cite{waterman_toward_2021} for maldaimonia; refer to the table notes for details.}
\label{tab:motivthemes}
\begin{tabularx}{\textwidth} {p{65pt}p{70pt}p{280pt}p{50pt}}
\hline
\makecell{Theme} & Sub-Theme & Description & Criteria \\
\hline
\makecell{Passivity} & Compliance & Encouraged or forced by external factors, especially the game design. & (3) \\
\multirow[t]{2}{*}{\makecell{Provocation}} & Revenge & Sought revenge due to a perceived wrong or wishing to ``do unto others.'' & (2), (4) \\
& Warmongering & Driven to carry out acts of violence and aggression. & (1), (2), (4) \\
\multirow[t]{3}{*}{\makecell{Mentalization}} & Escapism & Desired to escape the stresses and monotony of real life. & (1), (2), (4) \\
& Moral Hall Pass & Took advantage of the game context a means of doing forbidden or socially unacceptable things. & (1), (4) \\
& Extraordinarity & Desired to experience something extraordinary, something far outside the bounds of real life. & (1), (2), (3) \\
\multirow[t]{2}{*}{\makecell{Affective Drivers}} & Boredom & Sought a thrill due to boredom. & (1) \\
& Amusement & Seeking entertainment or an enjoyable time. & (1) \\
\multirow[t]{2}{*}{\makecell{Extrinsic Drivers}}& Desires & Sought extrinsic gratification, such as through acquiring money or goods, or status, or praise, or to win. & (1), (2), (3), (4) \\
& Influences & Inspired by external forces, such as word-of-mouth or reputation. & (1), (2) \\
\hline
\addlinespace
    \multicolumn{4}{@{}p{\dimexpr\linewidth}@{}}{\footnotesize (1) attaching positive valence to egocentric, destructive, and/or exploitative activities. (2) providing a basis of personal identity. (3) striving for excellence or mastery in these activities. (4) aligning these activities as acts of personal expressiveness.}
\end{tabularx}
\end{table*}

\begin{table*}
\caption{Maldaimonic game experiences framework from \citet{seaborn_meaningful_2023}. Criteria refer to the four criteria proposed by Waterman~\cite{waterman_toward_2021} for maldaimonia; refer to the table notes for details.}
\label{tab:malthemes}
\begin{tabularx}{\textwidth} {p{65pt}p{70pt}p{280pt}p{50pt}}
\hline
\makecell{Theme} & Sub-Theme & Description & Criteria \\
\hline
\multirow[t]{2}{*}{\makecell{Transgressions}} & Murder \& Mayhem   & Destructive acts, including killing, maiming, and harm, as well as destroying property and/or environments. & (1) \\
 & Chaos & Actions that lead to confusion, instability, or disorder, or have no point or impetus. & (1) \\
\multirow[t]{3}{*}{\makecell{Reflections}} & Rule Subversion & Getting away with morally and/or ethically deviant actions, including exploitation, stealing, looting, cheating, insults, and shady deals. & (1), (3) \\
& Hubris & Having pride and confidence, especially extreme pride and overconfidence. & (2), (3), (4) \\
& Vengeance & Reacting to perceived or actual slights with vengeful acts, often of escalating severity. & (1), (4) \\
\multirow[t]{3}{*}{\makecell{Feelings}} & Malight & Malicious delight (``malight'') directly linked to the maldaimonic experience. & (1) \\
& Power & Expressions of power and invincibility, and the successful use of force. & (3) \\
& Mood Shifts & Moods and affective states influence or are influenced by maldaimonia. & (2) \\
\multirow[t]{2}{*}{\makecell{Appreciations}} & Extrinsic Appetite & Satisfaction of needs and desires through external factors, such as rewards, praise, fame, and collecting material goods. & (3), (4) \\
& Aesthetics & Recognizing the visuals, sounds, animations, and other sensory features. & (4) \\
\hline
\addlinespace
    \multicolumn{4}{@{}p{\dimexpr\linewidth}@{}}{\footnotesize (1) attaching positive valence to egocentric, destructive, and/or exploitative activities. (2) providing a basis of personal identity. (3) striving for excellence or mastery in these activities. (4) aligning these activities as acts of personal expressiveness.}
\end{tabularx}
\end{table*}

\subsubsection{Quantitative Analysis}

We conducted inferential analyses, focusing on comparing the Japanese and US player measures and exploring the presence of relevant orientational, experiential, and game UX factors, as per the sub-RQs. We assessed the distribution of quantitative data encompassing Hedonic and Eudaimonic Motivations (HEMA), Player Experience (PXI), Positive and Negative Affect (PANAS-X), and Long-term Meaning Potential using the Shapiro-Wilk test. All were non-normal (refer to the Supplementary Materials). As such, we opted for the Mann-Whitney non-parametric test. We did not remove outliers as none were extreme: all lay within the data range for each measure. Also, the Mann-Whitney, as a non-parametric test, is robust against outliers~\cite{zimmerman1994note}. Although the presence of a large number of outliers may affect power, this was not the case for our data~\cite{zimmerman1994note}. We also employed Kendall's tau-b, Spearman's rho, and Pearson correlations (as the more common statistic, with a view to future meta-synthesis work) to explore the theorized connections between measures. 

Due to a technical glitch in the online survey for US participants, a few variables 
in the PANAS-X questionnaire were not recorded. These variables were ``upset'' ``distressed,'' ``active,'' ``inspired,'' ``interest,'' ``disgust,'' and ``dissatisfied.'' Consequently, we have excluded these variables for both countries in our analysis.

\section{Qualitative Findings}
We now report our findings, starting with the qualitative data describing the contexts of the experiences, the maldaimonic play motivations people had, and what maldaimonic experiences occurred. We then explore the quantitative side: correlates of maldaimonic UX among the measures captured. In each section, we compare our Japanese cohort to the US one (N=51) captured by \citet{seaborn_meaningful_2023} for our cross-cultural analysis of maldaimonic game UX.

\subsection{Where and Who in Maldaimonic UX (RQ1)}
\subsubsection{Gaming Context}
Respondents reported maldaimonic experiences in a variety of games: 44 unique games were mentioned by 59 participants. We use the game genre classification for games in Japan by NIHON KOGAKUIN\footnote{\url{https://www.neec.ac.jp/department/design/gamecreator/type/}}. As such, genres included: adventure (15, 25\%; e.g. Grand Theft Auto series, Resident Evil series, etc.), RPG (13, 22\%; e.g. Pocket Monster, Dragon Quest series), shooter (10, 17\%; e.g. Splatoon, call of duty, etc.), simulation (7 respondents, 12\%; e.g. SimCity), action (7 respondents, 12\%; e.g. Super Mario, Super Smash Bros.), racing (4 respondents, 7\%; e.g. Fall Guys), sandbox (Minecraft), and tabletop (Shogi). The platforms were PlayStation (22, 37\%), PC (13, 22\%) smartphones (7, 12\%), Nintendo Switch (5, 8\%), Nintendo DS (5, 6\%) and other (4, 6\%).

The variety of platforms and game genres echoes that of the US cohort. Notably, the top genres were action-adventure (US: 11, 23\%), RPG (US: 7, 15\%; Japan: 13, 22\%), and simulation (US: 5, 11\%; Japan: 7, 12\%). There were some differences: MMO was a top genre for US players (7, 15\%), while shooters (10, 17\%) and adventure (Japan: 15, 24\%) were top genres for Japanese players. Still, the spread of the numbers suggests this could be a sample size issue.

\subsubsection{Social Context (RQ1)}
The majority, 33 (56\%), had maldaimonic experiences alone, 14 (23\%) with two people, and 12 (20\%) with 2+ people.
In 14 cases (23\%), there were two players, and in 12 cases (20\%), there were multiple players in multiplayer mode. In terms of who, the majority played with known others: 15 (58\%) were with friends or family, nine (34\%) with strangers, and two (7\%) with a mix of friends, family and strangers. 

As with the US cohort, these patterns provide further support for distinguishing maldaimonic game UX from dark participation~\cite{kowert_dark_2020,kowert_toxicity_2022}. The Japan and US (29, 57\%) players characterized such events as individual experiences, in line with Waterman's criteria for maldaimonia as a construct of identity formation and expression \cite{waterman_toward_2021}. Moreover, when others were involved, they were known to the player, as was the case for the US cohort: 24 (44\%) were social, but all of these involved friends, family, or acquaintances, even if strangers were sometimes involved. From a dark participation perspective, this is unexpected, as bad actors tend to hide behind anonymity \cite{nitschinsk_disinhibiting_2022}. While some incidents could have been rare bad events, as we find below, this was largely not the case, suggesting a level of social acceptability that could relate to the ``magic circle'' of the game~\cite{huizinga2014homo,salen_rules_2004} or ``having fun being bad'' among friends.

\subsection{Why Play: Patterns of Maldaimonic Orientations and Motivations (RQ1)} \label{sec:whyplay}

Motivations are presented in~\autoref{tab:motivmalfactors}. 48 of 59 (81\%) participants provided an explanation of their maldaimonic play motivations. These ran the gamut offered by the framework. Most Japanese players were motivated to comply with the game rules. Others sought an escape, amusement, or took advantage of the game context to try something morally transgressive. Several were influenced by external factors, especially suggestions by others or knowledge of the game. For instance, respondents wrote about being curious after witnessing others do similar things in videos or hearing about it from friends. Some were seeking an extraordinary experience or desired to satisfy non-social extrinsic needs. A few cited boredom or were provoked into revenge or rampage by others.

Chi-squared tests found no statistically significant differences between Japanese and US players for any sub-themes.

In sum, Japanese players were motivated to play maldaimonically for a variety of reasons that were similar in nature and frequency to those of the US players.

\begin{table*}
\caption{Motivations for playing maldaimonically. JP n=48/59, US N=51.}
\label{tab:motivmalfactors}
\begin{tabularx}{\textwidth} {p{65pt}p{60pt}p{275pt}p{30pt}p{30pt}}
\hline
\makecell{Theme} & Sub-Theme & JP Example & US & JP \\
\hline
\makecell{Passivity} & Compliance & P41:\phantom{ }ゲームのルールに沿ってプレイした (played according to the rules of the game) & 18 (35\%) & 10 (21\%) \\
\multirow[t]{2}{*}{\makecell{Provocation}} & Revenge & P15:\phantom{ }同じことをやられたから、やり返したくてやった (they did it to me, so I wanted to do it back)& 5 (10\%)& 3 (6\%)\\
& Warmongering & P18:\phantom{ }ストーリーで気に入らないキャラクターだったとき (when you didn't like a character in a story)& 2 (4\%) & 1 (2\%)\\
\multirow[t]{3}{*}{\makecell{Mentalization}} & Escapism & P37:\phantom{ }職場でのストレス (stress at work) & 5 (10\%)& 7 (15\%)\\
& Moral Hall Pass & P68:\phantom{ }ヒマな中学生の悪ふざけを、誰にも叱られずに行えたからだと思います (I was able to do the mischievous pranks of a junior high school student without being scolded by anyone)& 4 (8\%)& 7 (15\%)\\
& Extraordinarity & P39:\phantom{ }無敵になり、破壊力が増すことが快感だった為です (because it was pleasurable to become invincible and more destructive) & 4 (8\%)& 5 (10\%)\\
\multirow[t]{2}{*}{\makecell{Affective Drivers}} & Boredom & P38:\phantom{ }普通に遊んでも面白くなくなってきて、よりスリリングな方法を選んだように思う (I think it became less interesting to play in a normal way and I chose a more thrilling way)& 3 (6\%)& 2 (4\%)\\
& Amusement & P61:\phantom{ }敵の倒れる動作や弱っていく過程が面白かったため (because the process of dropping and weakening enemies was interesting)& 7 (14\%)&8 (17\%) \\
\multirow[t]{2}{*}{\makecell{Extrinsic Drivers}}& Desires & P49:\phantom{ }課金せずに効率よく強くなる方法はないかといろいろ探った結果ギルドのルールを無視しようと思った (after exploring ways to get stronger without paying, I decided to ignore the guild rules)& 7 (14\%)& 4 (8\%) \\
& Influences & P23:\phantom{ }評判が良かったゲームだったため (the game had a good reputation)& 2 (4\%)& 7 (15\%) \\
\hline
\end{tabularx}
\end{table*}

\subsection{What It Is: Maldaimonic Patterns in the Player Experience (RQ1)} \label{sec:whatitis}
In their \emph{critical incident accounts}, Japanese players reported on encountering or carrying out a vast array of maldaimonic \emph{game experiences} (\autoref{tab:expmalfactors}), similar to US cohort~\cite{seaborn_meaningful_2023}.

A Chi-squared test comparing Japanese and US players found a statistically significant difference for Malight, \textit{$X^2$}(1, \textit{N}=110) = 24.30, \textit{p} < .001, suggesting US players derived a greater sense of malicious pleasure compared to Japanese players. US players also significantly more often characterized their experience in terms of Aesthetics, or the visuals, sounds, animations and other visceral aspects of the experience, \textit{$X^2$}(1, \textit{N}=110) = 4.65, \textit{p} = .031. No other differences were found.

As with the US cohort (8, 16\%), very few (10, 17\%) Japanese player accounts could be classified as dark participation, i.e., intent to do real harm to (real) others (not NPCs). Also, there was no difference found by country. These experiences were often implied to be good fun among friends. P15, for example, described:

\aptLtoX[graphic=no,type=html]{
\begin{quote}\ 
友達を倒して物資を奪ったり、家を破壊するイタズラしたり、死ぬ場所に誘導する罠を仕掛けたり
\hfill\break
Beating your friends and taking their supplies, pranking them to destroy their houses, setting traps to lead them to places where they will die, etc.
\end{quote}
}
{\begin{quote}
友達を倒して物資を奪ったり、家を破壊するイタズラしたり、死ぬ場所に誘導する罠を仕掛けたり\\
Beating your friends and taking their supplies, pranking them to destroy their houses, setting traps to lead them to places where they will die, etc.
\end{quote}}

Some Japanese players sought vengeance (P9, P13, P48), others wrought havoc (P54, P55), others cheated (P59), and still others wished to prank their friends (P15, P53). Not all accounts were frivolous. P29, for instance, expressed regret at their actions:\phantom{ }むなしさが残った (left me with a feeling of emptiness). Dark participation~\cite{kowert_dark_2020, quandt_dark_2018} may bleed into aspects of these maldaimonic accounts, but it appears to be a distinct facet of game experiences.

\begin{table*}
\caption{Maldaimonic game experiences found in the critical incident accounts. JP N=59, US N=51.}
\label{tab:expmalfactors}
\begin{tabularx}{\textwidth} {p{60pt}p{70pt}p{250pt}p{30pt}p{30pt}p{20pt}}
\hline
\makecell{Theme} & Sub-Theme & JP Example & US & JP & Sig. \\
\hline
\multirow[t]{2}{*}{\makecell{Transgressions}} & Murder \& Mayhem & P46:\phantom{ }ノコノコを踏んで倒していくこと (stepping on Koopas and knocking them down) & 40 (78\%) & 41 (69\%) &\\
 & Chaos & P26:\phantom{ }都内の一般道を暴走して楽しむ (I enjoyed going out of control on ordinary city roads) & 9 (18\%) & 14 (24\%) & \\
\multirow[t]{3}{*}{\makecell{Reflections}} & Rule Subversion & P15:\phantom{ }友達を倒して物資を奪ったり、家を破壊するイタズラしたり、死ぬ場所に誘導する罠を仕掛けたり (beating your friends and taking their supplies, pranking them to destroy their houses, setting traps to lead them to places where they will die, etc.)& 20 (39\%)& 17 (19\%) &\\
& Hubris & P16:\phantom{ }現実では出来ないことを行える背徳感 (a sense of immorality from doing things I can't do in reality)& 5 (10\%) & 6 (10\%) &\\
& Vengeance & P48:\phantom{ }自分が助かることしか考えていない人がいたので、みんなで協力するふりをしてキルを奪い続けたこと (there were people who only thought about saving themselves, so we pretended to work together and kept getting kills)& 3 (6\%)&2 (3\%) & \\
\multirow[t]{3}{*}{\makecell{Feelings}} & Malight & P19:\phantom{ }普段できない破壊行動で快感を感じた (I felt pleasure in destructive behaviour that I wouldn't normally be able to do) & 32 (63\%)& 10 (17\%) & ***\\
& Power & P39:\phantom{ }マリオが無敵になった際、ブロックを高速で多数壊せることに楽しみを感じました (when Mario became invincible, I enjoyed being able to break many blocks at high speed)& 27 (53\%)& 24 (41\%) &\\
& Mood Shifts & P66:\phantom{ }平和にチャットなどをしてるところをいきなり殺したりすることで普段実生活では味わえない気持ちよさがあったように思う (the unusual, unreal feeling when you suddenly kill people who are chatting peacefully) & 12 (24\%)& 22 (37\%) &\\
\multirow[t]{2}{*}{\makecell{Appreciations}} & Extrinsic Appetite & P64:\phantom{ }敵キャラクターを倒すことで経験値が得られ、自分のキャラクターが成長することが楽しかった (I enjoyed gaining experience and growing my character by defeating enemies)& 17 (33\%)& 14 (24\%) &\\
& Aesthetics & P68:\phantom{ }エンジン等の出力を最大にした車輌でオーバルコースの壁に向かってフルスロットルで横滑りさせながら走ると、1000km/hを超える速度が出て処理落ちが発生し紙芝居のようになったあと、本来走行できない壁の外を走ったり、コースの下へ無限に落ち続けたりした (I drove a vehicle, engine to the max, and skidded at full throttle towards the wall, the speed exceeded 1000 km/h and a bug happened, it became like a paper screen, I drove outside the wall and kept falling infinitely down in space)& 6 (12\%)&1 (2\%) & * \\
\hline
    \addlinespace
    \multicolumn{6}{l}{\footnotesize *\textit{p} < .05. ***\textit{p} < .001.}
\end{tabularx}
\end{table*}

\subsection{Theoretical Congruity: Mapping Onto Waterman's Four Criteria (RQ2)}

Participants described motivations and experiences that could be linked back to Waterman's four criteria for maldaimonia~\cite{waterman_toward_2021} by way of our thematic analyses. Results are presented in~\ref{tab:freqcriteria}. We have indicated where no criterion was found, as well as where 75\% were found, as a baseline level of understanding the extent of attributions. Since maldaimonia is orientational and experiential, in theory, we need to consider both together. Results indicate that the majority of accounts (39 or 81.3\%) satisfied the majority of Waterman's criteria. 43.8\% (or 21) accounts satisfied all criteria.

\begin{table}
\caption{Frequency of the four criteria for maldaimonia found in the motivation and experience themes.}
\label{tab:freqcriteria}
\begin{tabularx}{\linewidth} {p{25pt}p{25pt}p{25pt}p{25pt}p{25pt}p{25pt}p{25pt}}
\hline
\multirow[t]{2}{*}{\makecell{Criteria}} & \multicolumn{2}{l}{Motivations} & \multicolumn{2}{l}{Experiences} & \multicolumn{2}{l}{Both} \\
 & Count & \% & Count & \% & Count & \% \\
\hline
\makecell{None} & 3 & 6.3\% & 1& 1.4\% & 0& 0\% \\
\makecell{1} & 15 & 31.3\% & 9& 12.7\% & 2& 4.2\% \\
\makecell{2} & 11 & 22.9\% & 20& 28.2\% & 7& 14.6\% \\
\makecell{3} & 10 & 20.8\% & 21& 29.6\% & 18& 37.5\% \\
\makecell{4} & 9 & 18.8\% & 8& 11.3\% & 21& 43.8\% \\
\makecell{3 \& 4} & 19 & 39.6\% & 29& 40.8\% & 39& 81.3\% \\
\hline
\addlinespace
    \multicolumn{7}{@{}p{\dimexpr\linewidth}@{}}{\footnotesize (1) attaching positive valence to egocentric, destructive, and/or exploitative activities. (2) providing a basis of personal identity. (3) striving for excellence or mastery in these activities. (4) aligning these activities as acts of personal expressiveness.}
\end{tabularx}
\end{table}

\section{Quantitative Findings}
We now report the quantitative findings, measure by measure.

\subsection{Positive and Negative Affect (PANAS-X) (H1)} \label{sec:panas}
We evaluated the PANAS-X scores for each subscale: 1) General Dimension scale (Negative Affect, Positive Affect); 2) Basic Negative Emotion scale (Fear, Hostility, Guilt, Sadness); 3) Basic Positive Emotion scale (Joviality, Self-Assurance, Attentiveness); and 4) Other Affective States (Shyness, Fatigue, Serenity, Surprise). 

\begin{figure*}
 \centering
 \begin{subfigure}[b]{0.49\textwidth}
     \centering
     \includegraphics[width=\textwidth]{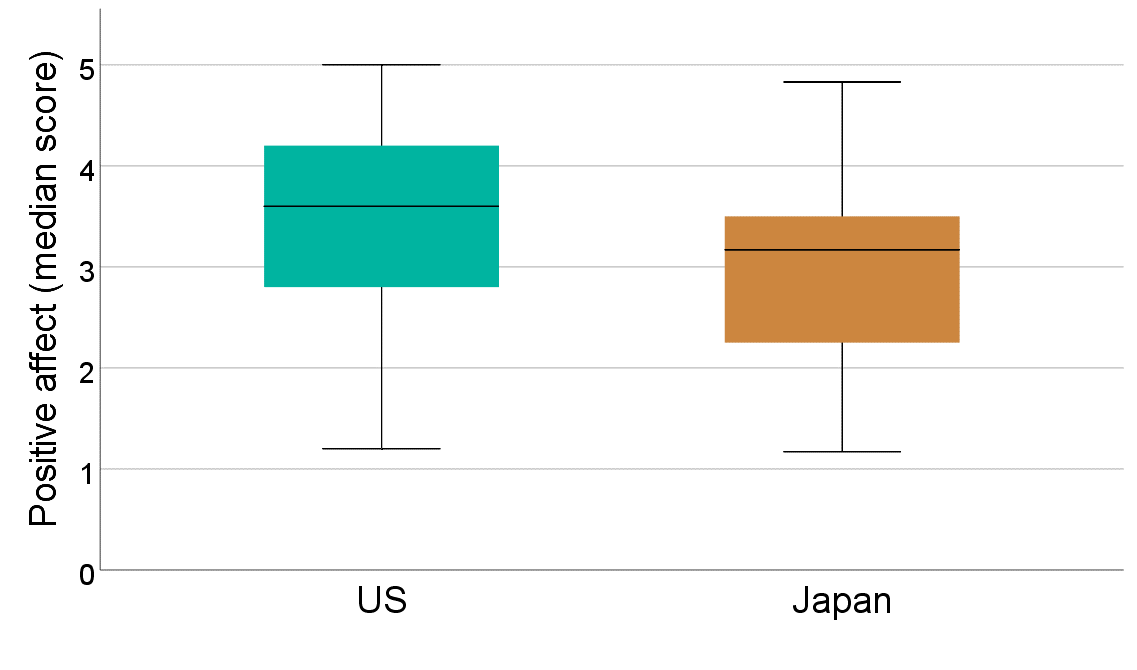}
     \caption{}
     \label{fig:positiveaffect}
 \end{subfigure}
 \begin{subfigure}[b]{0.49\textwidth}
     \centering
     \includegraphics[width=\textwidth]{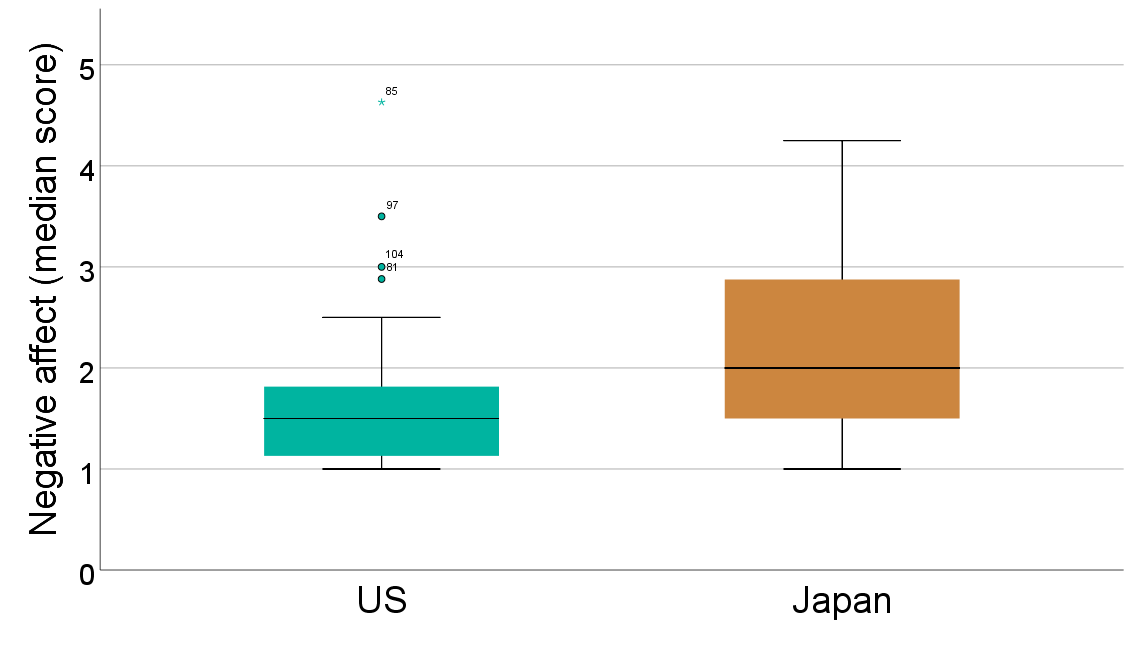}
     \caption{}
     \label{fig:negativeaffect}
 \end{subfigure} 
 \caption{Positive affect scores (left) were statistically significantly higher for US participants. Negative affect scores (right) were statistically significantly higher for Japanese participants.}
 \label{fig:affect}
 \Description{Positive affect scores (left) show higher significant scores for US participants. Negative affect scores show higher (right) statistically significant scores for Japanese participants.}
\end{figure*}

For the general dimension scale, the Mann-Whitney U score related to negative affect showed statistically significantly higher scores for the Japanese participants (MD=2.0) compared to the US participants (MD=1.6), \textit{U} = 917.5, \textit{r} = -.33, \textit{p} < .001. However, the results for positive affect showed statistically significantly higher scores for the US participants (MD=3.6) compared to the Japanese participants (MD=3.17), \textit{U} = 1050.5, \textit{r} = -.25, \textit{p} = .006 (\autoref{fig:affect}).

\begin{figure*}
 \centering 
 \begin{subfigure}[b]{0.49\textwidth}
     \centering
     \includegraphics[width=\textwidth]{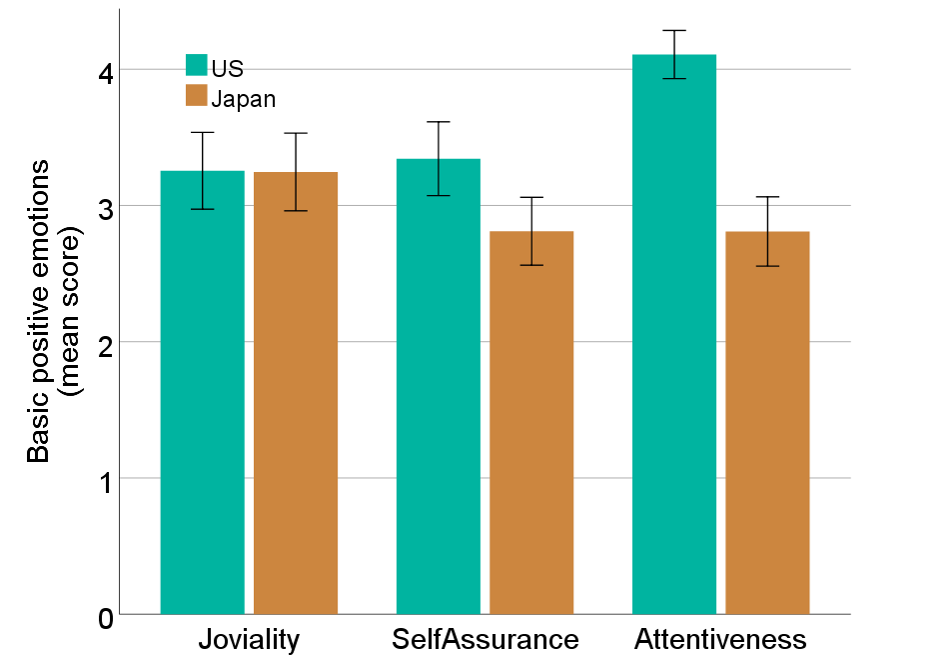}
     \caption{}
     \label{fig:basicpositiveall}
 \end{subfigure}
 \begin{subfigure}[b]{0.49\textwidth}
     \centering
     \includegraphics[width=\textwidth]{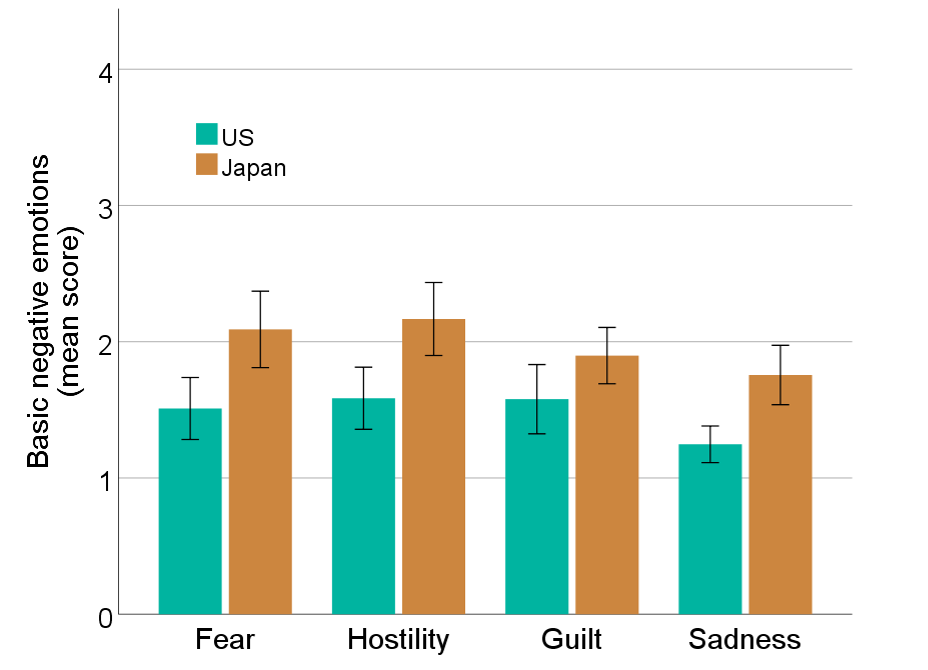}
     \caption{}
     \label{fig:basicnegative}
 \end{subfigure}  
 \begin{subfigure}[b]{0.49\textwidth}
     \centering
     \includegraphics[width=\textwidth]{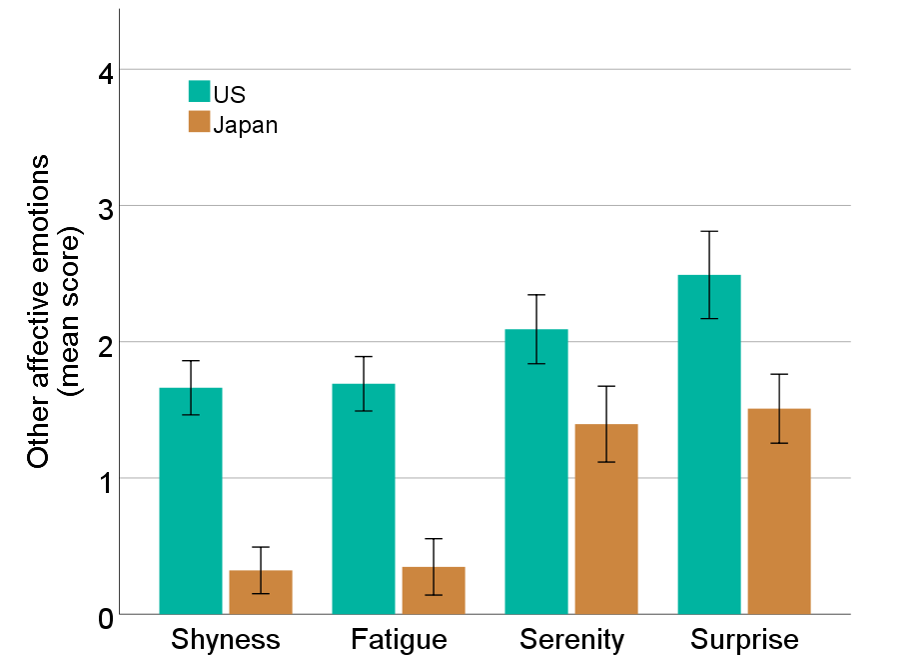}
     \caption{}
     \label{fig:otheraffective}
     \end{subfigure}
 \caption{Basic positive emotion scores related to self-assurance and attentiveness were statistically significantly higher for US participants (\autoref{fig:basicpositiveall}). All basic negative emotion scores were statistically statistically significantly for Japanese participants (\autoref{fig:basicnegative}). Other affective scores show higher values for US participants (\autoref{fig:otheraffective}).}
 \label{fig:posneg}
 \Description{Basic positive emotion scores for self-assurance and attentiveness show significantly higher scores for US participants. All basic negative emotion scores show statistically significant higher scores for Japanese participants. Other affective scores show higher values for US participants.}
\end{figure*}

The results of the other subscales are shown in \autoref{tab:panassubscales}. While there was no statistically significant difference for Joviality, all other factors showed a difference between the US and Japanese cohorts (\autoref{fig:basicpositiveall}). Japanese respondents expressed higher levels of basic negative emotions, including Fear, Hostility, Guilt, and Sadness (\autoref{fig:basicnegative}). Meanwhile, US respondents expressed higher levels of basic positive emotions, including Self-assurance and Attentiveness, and other affective states, including Shyness, Fatigue, Serenity, and Surprise (\autoref{fig:otheraffective}). In general, US respondents experienced higher levels of emotion, especially positive affect, while Japanese respondents experienced more negative affect.

In sum, \textbf{we can partially accept H1: that maldaimonic game experiences were often, but not always, or solely, positive}.

\begin{table*}
\caption{Statistical results for positive and negative affect via the three subscales of the PANAS-X.}
\label{tab:panassubscales}
\begin{tabularx}{.705\linewidth} {lllllll}
\hline
\makecell{Subscale} & Factor & \multicolumn{5} {l} {Statistics}\\
& & US MD & JP MD & \textit{U} & \textit{r} & \textit{p} \\
\hline
\multirow[t]{4}{*}{\makecell{Basic Negative Emotion Scales}} & Fear & 1.14& 1.83 & 992.5 & -0.29 & .002**\\
 & Hostility & 1.17& 2.0 & 947.5 & -.32 & .001** \\
 & Guilt & 1.25& 1.75 & 1061.5 & -.25 & .006** \\
 & Sadness & 1.0& 1.6 & 827.5 & -.4 & < .001*** \\ 
\multirow[t]{3}{*}{\makecell{Basic Positive Emotion Scales}} & Joviality & 3.3& 3.3 & 1481 & -.01 & .888 \\
 & Self-assurance & 3.5& 2.8 & 1012 & -.27 & .003** \\
 & Attentiveness & 4.0& 3.0 & 417.5 & -.62 & < .001*** \\ 
\multirow[t]{4}{*}{\makecell{Other Affective States}} & Shyness & 1.5&  0 & 264 & -.725 & < .001*** \\
 & Fatigue & 1.5& 0 & 280 & -.725 & < .001*** \\
 & Serenity & 2.0& 1.33 & 891 & -.35 & < .001*** \\
 & Surprise & 2.33& 1.33 & 814 & -.39 & < .001*** \\          
\hline
    \addlinespace
    \multicolumn{7}{@{}p{\dimexpr\linewidth}@{}}{\footnotesize *\textit{p} < .05. **\textit{p} < .01. ***\textit{p} < .001.}
\end{tabularx}
\end{table*}

\subsection{Hedonic and Eudaimonic Motivations (HEMA) (H2, H3)} \label{sec:hema}

First, we ran Kendall's tau-b and Spearman's rho correlations to determine the expected association between hedonia and eudaimonia. Japanese participants exhibited a moderate, statistically significant positive correlation (Kendall's tau-b: \textit{$\tau$} = .369, \textit{n} = 59, \textit{p < .001}; Spearman's rho, \textit{$r(57)$} = .498, \textit{p < .001}). Thus, \textbf{we can confirm the theorized relationship between eudaimonia and hedonia within the context of maldaimonia (H3)}.

\begin{figure*}
 \centering
 \begin{subfigure}[b]{0.49\textwidth}
     \centering
     \includegraphics[width=\textwidth]{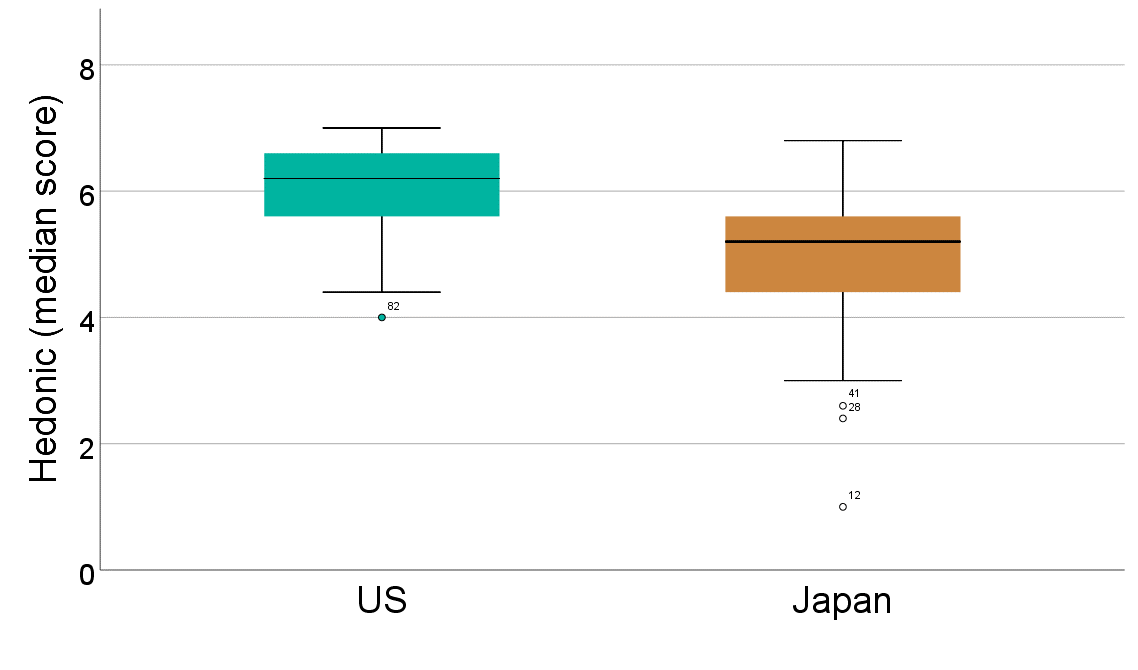}
     \caption{}
     \label{fig:hedonic}
 \end{subfigure}
 \begin{subfigure}[b]{0.49\textwidth}
     \centering
     \includegraphics[width=\textwidth]{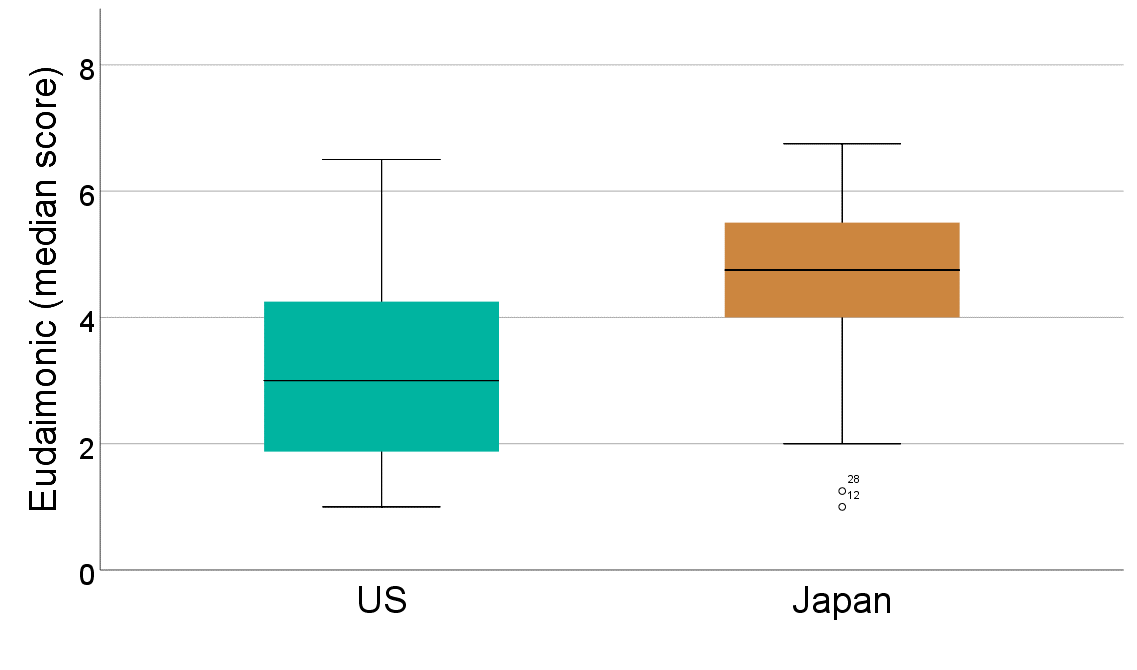}
     \caption{}
     \label{fig:eudaimonic}
 \end{subfigure} 
 \caption{Hedonic scores (left) were statistically significantly higher for US participants. Eudaimonic scores (right) were statistically significantly higher for Japanese participants.}
 \label{fig:hideud}
 \Description{Hedonic and Eudaimonic scores showing statistically significant differences between the US and Japanese participants.}
\end{figure*}

Next, we ran a Mann-Whitney U test to determine any differences between the US and Japanese participants regarding their hedonic and eudaimonic motivations (\autoref{fig:hideud}). The hedonic scores of the US participants (MD=6.2) were greater than the Japanese participants (MD=5.2), \textit{U} = 479, \textit{r} = -.58, \textit{p} < .001.
However, the US participants (MD=3.0) had lower eudaimonic scores compared to Japanese participants (MD=4.7), \textit{U} = 715.5, \textit{r} = -.45 \textit{p} < .001. Thus, \textbf{we can confirm that players were meaning-oriented during maldaimonic play, especially Japanese players (H2)}.

\subsection{Player Experience (PXI) (H2, H5)} \label{sec:pxi}
We conducted a Mann-Whitney U test to evaluate whether there were differences between the play experiences of US and Japanese participants.
Participants from the US (MD=1.87) and Japan (MD=1.93) did not significantly differ in terms of psychosocial consequences, \textit{U} = 1210, \textit{r} = -.08, \textit{p} = .07.  
Similarly, US players (MD=1.87) did not seem to differ from Japanese players (MD=1.93) in terms of functional consequences, \textit{U} = 1362, \textit{r} = -.16, \textit{p} = .39. The generally high, positive scores for both dimensions \textbf{provides additional support for H2 (meaning) and confirms H5 (good player UX)}.

\subsection{Long-term Meaning Potential (H4)} \label{sec:longterm}
We ran Kendall's tau-b, Spearman's rho, 
and Pearson correlation tests to determine the relationship between eudaimonic (MD = 4), hedonic (MD=5.6), and long-term meaning potential scores (MD = 3.5). The Kendall's correlation test showed a weak, positive correlation for Japanese participants between their eudaimonic scores and long-term meaning potential, which was statistically significant (\textit{$\tau$} = .195, \textit{n} = 59, \textit{p} = .048). Similarly, the results for Spearman's rho showed a weak, positive correlation between eudaimonic scores and long-term meaning potential, which was statistically significant (\textit{$r(57)$} = .267, \textit{p} = .041).
However, the results of the Pearson correlation were not statistically significant (\textit{r} = .202, \textit{n} = 59, \textit{p} = .125). As predicted, no relationship was found with hedonia. No statistically significant results were found for the US cohort, either. While \textbf{the theorized relationship between long-term meaning potential and eudaimonia in the context of maldaimonia appears to exist for Japanese players (H4)}, in light of the above, we must interpret these results cautiously~\cite{khamis2008measures}.

\section{Discussion}
Maldaimonia manifests within the realm of Japanese gaming in ways similar but sometimes different to that which US players experience. Here, we discuss the findings, reflect on theory and methodology, and trace out paths for future work.

\subsection{The Nature of Maldaimonic Play in Japan (RQ2)}
Player accounts show that maldaimonic game UX is an experiential and orientational construct engendered through a \textbf{\textit{variety of motivations and experiences}} linked to Waterman's four criteria~\cite{waterman_toward_2021} and found across an \textbf{\textit{array of games and gaming contexts}}.
Furthermore, it exhibits connections, both anticipated and unexpected, with its conceptual counterpart, \textbf{\textit{eudaimonia}} (H2), eudaimonia's correlate, \textbf{\textit{hedonia}} (H3), and various aspects of the gaming experience, including \textbf{\textit{player experience}} (H5), notably functional and psychosocial needs, \textbf{\textit{positive emotional states}} (H1), and \textbf{\textit{long-term meaning potential}} (H4).

Few accounts had the combined purposive and social nature of cruel play~\cite{sutton-smith_ambiguity_2001} or dark participation in games~\cite{kowert_dark_2020,kowert_toxicity_2022}. Most were found in single-player contexts (33 or 56\%) and social contexts with known others (17 or 65\%). Revenge and vengeance were rare, while warmongering and acts of murder and mayhem were mostly unconnected to real people (refer to Tables~\ref{tab:motivmalfactors} and \ref{tab:expmalfactors}). Similarly, no accounts connoted the conscious rebellion that \citet{aarseth2014fought} articulates for transgressive play. The themes of moral hall pass (7 or 15\%), chaos (14 or 24\%), and rule subversion (17 or 19\%) speak to rule-breaking within the game. Yet, these do relate to transgression as characterized by \citet{mortensen_paradox_2020}. Within each of these modes of maldaimonia, we may find the two varieties of transgression that they delineate: \emph{extraludic} transgressions related to norms in society and \emph{intraludic} ones tied to the game itself. As they note, there can be a fuzzy line between the two, since games can be modelled on the real world or involve others who bring in expectations around extra-game social norms.
At the same time, the presence of the compliance theme (10 or 21\%) suggests the opposite of rebellion against the game. This points to the notion of dark play~\cite{mortensen_dark_2015,mortensen_paradox_2020}, where the game design itself stimulated maldaimonic play. While out of scope for this paper, future work could compare and contrast maldaimonic experiences resulting from games designed and not designed to be provocative. In sum, maldaimonic game UX represents \textbf{\textit{meaningful solo player experiences that enable immersive self-expression through egocentric, exploitative, and/or destructive acts}} or multiplayer scenarios where the transgressive elements tend to \textbf{\textit{align with social norms}} among peers and within social networks.

This juxtaposition to dark and transgressive play prompts us to consider maldaimonia against the ``magic circle'' of the game. First conceptualized by \citet{huizinga2014homo} and later popularized by \citet{salen_rules_2004} for digital games, the \emph{magic circle} refers to the artificial space that we, the players, voluntarily cross into when embarking on play journeys. Here, the rules we are familiar with may be absent or reconfigured. Huizinga characterizes this space as unimpeded by ``moral duty''~\cite[p.~8]{huizinga2014homo} and without ``moral function''~\cite[p.~6]{huizinga2014homo}. Yet, we may find a contradiction when meaning and ethics are considered together that would elude investigations on eudaimonic engagement, which is ethically positive. Maldaimonic engagement renders meaning conspicuous through negative moral weight. Player accounts referencing the real world and especially defending maldaimonic pursuits reflect this. Whether maldaimonia creates a new and murky liminal space between real world consequences and the unreality of play \emph{or} renders the magic circle obsolete is a question we leave to future philosophers of play. Still, multiplayer contexts may present a clearer picture. Disruption of the magic circle for \emph{other players} that is linked to a given player's maldaimonic excursions could draw the line between permissible maldaimonia and cruel play~\cite{sutton-smith_ambiguity_2001} or dark participation~\cite{kowert_dark_2020}.

Critical incidents were reported across a range of game genres and specific titles, indicating that maldaimonic game UX is not tied exclusively to particular types of games, even violent ones, contrary to Waterman's contemplations~\cite{waterman_toward_2021} and the dark play literatures. Indeed, respondents provided an array of malightful but non-violent acts and orientations that point to need satisfaction through extrinsic motivation~\cite{ryan_intrinsic_2000}, a key trajectory for future study.

One noteworthy deviation from the findings for the US cohort and the requirements proposed by \citet{waterman_toward_2021} related to \textit{positive} affect. This and several other nuanced results found on comparing the US and Japanese cohorts raise the possibility of cultural differences, to which we turn next.

\subsection{Cultural Congruities and Caveats: The Case of Japan vs. the US (RQ1)}
The differences for hedonia, eudaimonia, long-term meaning potential, and facets of affect highlight a current of \textbf{\textit{meaningful play}} in the Japanese cohort. Likewise, the focus on experiences of \textbf{\textit{malight}}, or \textbf{\textit{malicious delight}}, by the US cohort, points to a pattern of short-term maldaimonic satisfaction. Taken together, maldaimonic game UX could be \textbf{\textit{culturally-sensitive}}, driven to a greater degree by hedonic orientations for US players while fueled by eudaimonia for those in Japan. These implications should be considered when developing and validating a measure of maldaimonia.

Three dimensions from Hofstede's model~\cite{hofstede2001culture} may help explain some of these differences. Japan has generally been characterized as a \textbf{\textit{collectivist}} culture, while the States has been cast as individualistic~\cite{minkov2017revision}. The higher rates of negative emotions alongside qualitative insights, such as being inspired by external forces to try out maldaimonic acts, points to a more collectivist spirit, one that recognizes and engages with others. The US has also been characterized as indulgent and Japan a culture of \textbf{\textit{restraint}}~\cite{bergiel2012revisiting}. This touches on the divide between these cohorts in terms of short-term hedonic pursuits for the US and long-term meaning potential and eudaimonia for Japan. Finally, this last difference 
also matches greater levels of the Hofstedian \textbf{\textit{long-term orientation}} value found in Japan~\cite{bergiel2012revisiting}. These are admittedly broad strokes, patterns that should be critically explored with further mixed methods work.

We can summarize the core \textbf{\textit{similarities}} in maldaimonic game UX between the US and Japanese cohorts as follows:
\vspace{3.25pt}

\begin{itemize}
    \item A similar range and degree of motivations (refer to \ref{sec:whyplay}) and experiences (refer to \ref{sec:whatitis}), with caveats
    \item A similar degree of functional and psychosocial consequences in maldaimonic play (refer to \ref{sec:pxi})
    \item A similar range of affective states (refer to \ref{sec:panas}), with caveats
\end{itemize}

\vfill\null
We can then summarize the core \textbf{\textit{differences}} as follows:

\begin{itemize}
    \item Japanese players experienced comparatively less malight, or malicious delight, and associated their experience less with aesthetics than US players (refer to \ref{sec:whatitis})
    \item Japanese players were driven to a greater degree by eudaimonic orientations and, weakly, long-term meaning potential (refer to \ref{sec:hema} and \ref{sec:longterm}), while US players pursued hedonia to a greater degree (refer to \ref{sec:hema})
    \item US players experienced higher affective states overall and especially greater positive affect, while Japanese players experienced a higher degree of negative affect (refer to \ref{sec:panas})
\end{itemize}

\subsection{Methodology and Magic Circles}
Maldaimonia is largely viewed as taboo. Social acceptability biases, reputation management, observer effects, self- and other-deception ... these and more are potential barriers for research \cite{waterman_toward_2021}, at least in terms of accuracy. Yet, the magic circle of the game medium appears to have been effective at nullifying these concerns. It is likely that some respondents held back, but the bold and unadulterated nature of the critical incident accounts is comforting. 
Still, though rare, we should heed the accounts that involved maldaimonic behaviour against other (human) players, as this could signpost antisocial and maladaptive personalities \cite{gonzalez_relationship_2018, greitemeyer_are_2019}. 
Waterman also questioned whether the game medium itself is a viable arena or simply a mode of symbolic expression. 
Yet, game experiences can be real for people. People can and do get attached to NPCs \cite{coulson_real_2012} and develop close ties with non/anonymous others \cite{barnett_virtually_2010}, including in violent games \cite{frostling-henningsson_first-person_2009}. 
Games are a facet of our multifaceted lives, and maldaimonia may simply be one variety of the gaming facet. 

Standardized methods of evaluating and/or measuring maldaimonia will also be needed. \citet{waterman_toward_2021} proposed the development of a two-factor instrument, including personal expressiveness and ethicality scales. Schadenfreude, translated from the German as ``malicious delight,'' has been included in previous game instruments~\cite{de_kort_digital_2007}; items could be extracted and expanded upon to develop a comprehensive scale. The present work contributes two thematic frameworks on maldaimonic play motivations and experiences, expansions of the original frameworks by~\citet{seaborn_meaningful_2023}. 
These frameworks should be tested in future work and could seed the creation of a quantitative instrument.

We will also need to revisit the maldaimonic game UX frameworks~\cite{seaborn_meaningful_2023} when analyzing experiential accounts---and not just for cross-cultural work. While participants were asked to describe a maldaimonic experience and their motivations in line with Waterman's proposed four criteria for maldaimonia~\cite{waterman_toward_2021}, not all descriptions could be analyzed in a way that at least one sub-theme covering the criteria applied. Specifically, 56.3\% of accounts did not meet all four criteria and 33.9\% did not meet three. This does not necessarily mean that the thematic frameworks are unrepresentative of maldaimonia. Accounts were not substantial, with most being one-liners of an average 33 characters for motivations and 40 for experiences. Prompting a thick description may be necessary.
We recognize that this may also be a feature of our methods, i.e., use of an online survey. As such, we encourage follow-up work using other methods suitable for critical incident techniques, i.e., face-to-face interviews~\cite{woolsey_critical_1986}, as well as force a minimum word count when using online tools. We also raise the possibility of a scoping issue: reports on self-actions vs. those on the actions of others. Respondents on the receiving end do not necessarily know and may be hesitant to make strong claims about the motivations of others. Also, those carrying out maldaimonic initiatives may not be willing to admit their true feelings and motivations, even anonymously or with respect to solo play, due to the aforementioned social biases. Future work can separate those who willingly partake in maldaimonic experiences and those who mete it out.

\subsection{From Games to Maldaimonic UX}
The concept of maldaimonia that we have explored here is tied to the gaming context. Features such as malight and aesthetics are certainly functions of the game experience. However, people surely have maldaimonic experiences with other forms of interactive systems, environments, agents, and interfaces. We have all experienced yelling at our computer screens in frustration, even if we might not speak about it. Yet, we may be more open about these experiences if there is a way to do so. For example, through a ``Frustrometer'' or ``Squeezemouse,'' modes of interaction clearly marked as permissive of venting one's frustration, have been found effective~\cite{reynolds_sensing_2001}. Future work that moves maldaimonic game UX into a more general maldaimonic UX for HCI experiences will be challenging. There may be greater viability through non-invasive or even deceptive research approaches. This is not unprecedented; Wizard of Oz~\cite{dahlback_wizard_1993}, for instance, is a widely used and accepted deceptive approach to simulating in-progress prototypes, where a person pretends to be the computer, unbeknownst to the user. We can also take a cue from the distinction between maldaimonic game UX and dark participation: focusing on experiences with computers that do not involve other people. Despite these open questions and challenges, examining UX from a maldaimonic perspective may be fruitful for better understanding people's orientations towards and experiences with technology. Maldaimonia in games may be special or transferable. Careful and clever research designs will be needed to explore whether this is the case.

\subsection{Limitations}
Procedures that involve post hoc recall are limited~\cite{huta_pursuing_2010,kim_pleasure_2014,mekler_momentary_2016}. However, we cannot predict when maldaimonic experiences will occur in advance, so it is difficult to get around this challenge. The descriptions provided were not as substantial as prompted, potentially an issue with the online format and/or the novelty of the maldaimonia concept. Future work should expand on our methods with multi-tiered prompts, word limits, and supplementary approaches, such as interviews. The inductive analysis was carried out by a non-Japanese researcher, albeit one experienced in this context and with this method, and with input from Japanese researchers. Future work should be conducted to confirm the emergent themes derived from the Japanese data set and possibly identify new ones using only inductive approaches and stronger involvement of Japanese researchers. As mentioned, a Google Form error prevented us from capturing all factors in the PANAS-X. This could have affected the power of the instrument, especially the stability of each construct and its ability to capture the affective experience. Similarly, a minimum character count was not used, where its use could have bolstered the richness of the accounts. While the results nonetheless appear solid, subsequent work should correct these methodological issues. Ideally, future work will involve the use of a validated instrument for quantitative measurements of maldaimonic orientations and experiences. Finally, we conducted an exploratory analysis because of the novelty of maldaimonia. Future work will need to develop hypotheses and confirm the inferential statistics results for Japan, the US, and other cultural groups.

\section{Conclusion}
Maldaimonic UX appears be a cross-cultural orientational and experiential construct in game contexts, adding to the existing literatures on ``dark'' and ``transgressive'' play. 
We have shown how Japanese player's experiences of maldaimonia largely map onto those of a US cohort---and where they do not. We offer more robust empirical evidence of maldaimonia in gameplay through our qualitative analyses and relevant instrumental measures. Now that we know maldaimonia crosses cultural boundaries, we can act on the implications of its existence for other contexts: cultures but also HCI spaces and sites of play. The critical incident reports shared by participants detail maldaimonic encounters and drivers within games. These insights can be applied in the development and examination of user experiences in gaming and beyond. Notably, we can move towards scale development in a culturally-sensitive way.

\begin{acks}
This work was funded by the department and in part by the World Research Hub (WRH) Program of the International Research Frontiers Initiative (IRFI), Tokyo Institute of Technology. Our gratitude to Peter Pennefather for years of engaging discussion on eudaimonia and more recent lively debates on maldaimonia. We used ChatGPT to rewrite parts of the related work and discussion for distinction from the earlier work-in-progress paper~\cite{seaborn_meaningful_2023}. Katie Seaborn conscientiously dissents to in-person participation at CHI this year; read their positionality statement here: \url{https://bit.ly/chi24statement}
\end{acks}

\bibliographystyle{ACM-Reference-Format}
\balance
\bibliography{refs}

\end{CJK}
\end{document}